# SolRPDS: A Dataset for Analyzing Rug Pulls in Solana Decentralized Finance

Abdulrahman Alhaidari*
Informatics and Networked Systems
University of Pittsburgh
Pittsburgh, PA, USA
aba70@pitt.edu

Bhavani Kalal*
Computer Science and Engineering
Indian Institute of Technology
Kharagpur, India
monu223@kgpian.iitkgp.ac.in

Balaji Palanisamy
Informatics and Networked Systems
University of Pittsburgh
Pittsburgh, PA, USA
bpalan@pitt.edu

Shamik Sural
Computer Science and Engineering
Indian Institute of Technology
Kharagpur, India
shamik@cse.iitkgp.ac.in

## Abstract

Rug pulls in Solana have caused significant damage to users interacting with Decentralized Finance (DeFi). A rug pull occurs when developers exploit users' trust and drain liquidity from token pools on Decentralized Exchanges (DEXs), leaving users with worthless tokens. Although rug pulls in Ethereum and Binance Smart Chain (BSC) have gained attention recently, analysis of rug pulls in Solana remains largely under-explored. In this paper, we introduce SolRPDS (Solana Rug Pull Dataset), the first public rug pull dataset derived from Solana's transactions. We examine approximately four years of DeFi data (2021-2024) that covers suspected and confirmed tokens exhibiting rug pull patterns. The dataset, derived from 3.69 billion transactions, consists of 62,895 suspicious liquidity pools. The data is annotated for inactivity states, which is a key indicator, and includes several detailed liquidity activities such as additions, removals, and last interaction as well as other attributes such as inactivity periods and withdrawn token amounts, to help identify suspicious behavior. Our preliminary analysis reveals clear distinctions between legitimate and fraudulent liquidity pools and we found that 22,195 tokens in the dataset exhibit rug pull patterns during the examined period. SolRPDS can support a wide range of future research on rug pulls including the development of data-driven and heuristic-based solutions for real-time rug pull detection and mitigation.

---

*Both authors contributed equally.

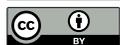

## 1 INTRODUCTION

Decentralized exchanges (DEXs) attract many users as they avoid intermediaries in peer-to-peer transactions [27]. However, DEXs, such as Jupiter on Solana, are less regulated compared to centralized cryptocurrency exchanges (CEXs), such as Coinbase [5]. As of November 2024, the total value locked (TVL) in decentralized finance (DeFi) exceeds $108.94 billion with a daily volume of $15.30 billion[11]. This makes it a lucrative environment for malicious activities. The anonymity of blockchain users complicates the regulation of DEXs, which creates opportunities for fraudulent schemes like rug pulls [17]. A rug pull occurs when malicious project developers promise high profits from joining a project. Once they achieve their target liquidity (assets stored in smart contracts) from users' deposits, they drain the liquidity pool and hide their traces, such as removing their social media presence and the project website [7].

Solana is a new blockchain launched in 2020 [28]. While rug pulls on Ethereum and Binance Smart Chain (BSC) have been studied recently [7], explorations focusing on Solana are limited despite the growing number of users and an increase in the number of reported rug pull instances. One of the reasons is the limited data available to support the understanding and development of new methods for mitigating them [25]. A key reason for the increasing number of incidents of rug pulls in Solana is its low gas fees. The average gas cost in Solana is $0.00025 per transaction. It also has high transaction throughput, processing up to 65,000 transactions per second (TPS) [28]. Users are increasing, and they become a target for malicious developers since the cost of creating tokens is negligible. Platforms like pump.fun[1] exacerbate this issue, as

---
[1]https://pump.fun/

Table 1: SolRPDS attributes and data types

| Attribute | Data Type | Description |
| --- | --- | --- |
| LIQUIDITY POOL ADDRESS | String | Address of the liquidity pool. |
| MINT | String | Token involved in the pool. |
| TOTAL ADDED LIQUIDITY | Float | Total liquidity added to the pool. |
| TOTAL REMOVED LIQUIDITY | Float | Total liquidity removed from the pool. |
| NUMBER OF LIQUIDITY ADDS | Integer | Number of liquidity addition transactions. |
| NUMBER OF LIQUIDITY REMOVES | Integer | Number of liquidity removal transactions. |
| ADD TO REMOVE RATIO | Float | Ratio of adds to removes. |
| FIRST POOL ACTIVITY TIMESTAMP | Timestamp | Timestamp of the first activity in the pool. |
| LAST POOL ACTIVITY TIMESTAMP | Timestamp | Timestamp of the last activity in the pool. |
| LAST SWAP TIMESTAMP | Timestamp | Timestamp of the last transaction. |
| LAST SWAP TRANSACTION ID | String | Transaction hash of the last swap. |
| INACTIVITY STATUS | Boolean | Indicates pool inactivity. |

they allow free token creation and enable developers to retain large portions of the token supply (the maximum number of tokens that can be traded), which increases the risk of rug pull.

In this paper, we introduce SolRPDS, the first public dataset for studying rug pulls on Solana. We derived potential cases from different DEXs from blockchain data over the time duration from February 2021 to November 2024. SolRPDS contains both suspected and confirmed rug pulls. Suspected cases are those where a token pool shows patterns such as rapid liquidity withdrawals or a significant decline in trading volume after liquidity removal [7], but users are still trading it. Confirmed instances are those where the liquidity was removed at once or gradually, and the token became inactive, which signifies users' realization of rug pull. In our dataset, the attributes mainly focus on capturing potential rug pulls and the attribute set includes token inactivity states, liquidity changes, last liquidity action, and last token swaps.

The main contributions of this paper are summarized below.

- Our dataset, SolRPDS, to the best of our knowledge, is the first public rug pull dataset for Solana that can be used for developing and evaluating new techniques including pattern identification for rug pull activities. SolRPDS is publicly available at: *https://github.com/DeFiLabX/SolRPDS*.
- SolRPDS simplifies the rug pull data retrieval from the blockchain and covers nearly four years of data (February 2021 to November 2024). The total liquidity pool actions examined are 278,099,218 and we have investigated 3,418,359,420 token swaps in 3,696,458,638 transactions for designing SolRPDS.
- We derived multiple attributes for identifying rug pull patterns, such as total liquidity removals, add-to-remove ratios, and inactivity periods based on the raw attributes directly from the blockchain. This aims to contribute to supporting future research for testing and evaluating new methods to mitigate rug pulls in Solana.
- In addition to derived attributes, we also have annotated tokens using the liquidity pool activity status based on the last DEXs interaction, which is a signal for rug pulls.
- We train and test six machine learning classifiers to show the utility of the data for future use in data-driven techniques.

## 2 RELATED WORK

Smart contracts in blockchains autonomously enforce agreements [29], but their logic flaws often expose them to attacks [19]. They can be used to implant vulnerabilities to enable attacks, such as a backdoor for rug pull. Studies like [3] have identified patterns in Ethereum Ponzi schemes for automated fraud detection, and [23] extended this work using network analysis to uncover fraudulent DeFi projects. However, rug pull investigations, especially in Solana, are still under-explored.

Existing Rug pull detection studies have primarily focused on Ethereum and Binance Smart Chain (BSC). In [2], for instance, authors have analyzed sudden liquidity removals to detect rug pull patterns. On the other hand, [26] identified fraudulent ERC20 token patterns. A notable work by [7] found that 60% of rug pull tokens on Ethereum and BSC last only a day, which they refer to as 1-day-token. They found that the liquidity withdrawals collectively have generated $240 million in profits.

Most rug pull research has focused on Ethereum and BSC because of their dominance in DeFi, such as [7, 21, 25]. Other blockchains, namely Solana, have been less investigated. There are no labeled datasets specific to Solana, to aid the development of detection or mitigation methods for rug pull behaviors. Because Solana lacks a public benchmark dataset, researchers face a challenge in validating mitigation mechanisms. We address this gap by introducing an extensively derived rug pull dataset for Solana, which is expected to help understand rug pulls in more detail and help create a safer DeFi ecosystem.

## 3 BACKGROUND

Blockchains facilitate the development of decentralized applications (dApps) [6] that run autonomously on blockchains in which transactions are recorded immutably. This enables secure and transparent transactions without the need for intermediaries [16]. Despite these properties, the blockchain ecosystem has new vulnerabilities that attackers exploit. Research on blockchain security has investigated widespread attacks such as phishing, Sybil attacks, smart contract exploits, and rug pulls [7, 21, 25]. In particular, rug pulls exploit the liquidity pools provided by users in decentralized exchanges (DEXs) to drain token value from the liquidity pool. Liquidity pools are token reserves in smart contracts that enable decentralized trading and lending by providing liquidity[13].

**Decentralized Finance (DeFi).** The growth of DeFi has introduced new risks, as financial services use smart contracts and liquidity pools that are vulnerable to exploitation and fraud [20]. DEXs are cryptocurrency exchanges that enable peer-to-peer trading of tokens. DEXs differ from CEXs, which act as custodians of user funds. DEXs allow users to maintain complete control over their assets as they interact with smart contracts and enable decentralization [17]. There are many popular DEXs such as Uniswap on Ethereum, PancakeSwap on BSC, and Jupiter on Solana. All of them rely on automated market maker (AMM) mechanisms, which use liquidity pools instead of order books to facilitate trading [27].

**DeFi and Rug Pulls:** Rug pulls in DeFi platforms have been increasing, where developers or liquidity providers withdraw assets from a pool that leaves investors with valueless tokens [18]. Studies such as [15, 22] discuss the increasing number of incidents of such fraud in DeFi and stress the need for creating robust detection mechanisms as the current methods are insufficient to protect investors. This is especially important for blockchain users, where low gas fees drive user growth. For instance, Solana, a low gas fee blockchain, recorded over 6.6 million active addresses in November 2024 [1]. However, this growth also makes Solana a target for rug pulls among other malicious activities.

**Solana** Solana has a unique architecture that does not rely on the Ethereum Virtual Machine (EVM). It has a custom runtime environment for smart contracts. Unlike other blockchains such as Ethereum, Avalanche, and Binance Smart Chain (BSC), which are EVM-compatible and share Ethereum-based functionality, Solana's unique design enables it to have a higher throughput and efficiency by using its own consensus mechanism. Solana runs on dual consensus mechanisms, namely Proof of History (PoH) combined with Proof of Stake (PoS) [28]. As Solana is known for its high performance and low transaction fees, it has become popular in the DeFi space [10]. However, its low gas fees and transaction speed create challenges for fraud detection [26] [3]. While rug pulls in other blockchains like Ethereum and Binance Smart Chain have been extensively studied, rug pull detection research specific to Solana still remains limited.

**Existing Datasets and Limitations.** There are a few archived blockchain data such as [14] that provide Solana blocks, accounts, and transactions, however, these are generic data. Downloading the data is challenging as it is cumbersome, which makes offline processing impractical for most current systems. These datasets also are not designed for security analysis, such as rug pull. In rug pull, some key features like the number of liquidity additions or removals, mint authority status, and inactivity signals are absent, making detection methods difficult to develop. To the best of our knowledge, there is no publicly available dataset tailored for rug pull analysis on Solana. This gap is emphasized by [25], which calls for annotated datasets for rug pull detection on Ethereum and BSC. A similar dataset for Solana is also essential to address rug pull and support the development of targeted solutions for mitigating fraudulent behaviors on Solana. While [7] analyzes over 1.3M tokens and 1M pools on Ethereum and BSC, identifying 272,349 rug pulls, their dataset is not publicly available. These platforms also benefit from older, standardized EVM infrastructure. In contrast, Solana was launched in late 2020 and it lacks such standards. Our dataset, tailored to Solana, is publicly available and fills this gap.

## 4 SolRPDS

The SolRPDS dataset captures rug pull incidents from transactions, liquidity pool, and token activity data over approximately a four-year period from February 12, 2021, to November 1, 2024. We analyzed Solana blockchain data using multiple sources, including [4, 8, 9], with Flipside [9] being the most suitable. Flipside provides a suite of analytical tools to examine real-time and historical data over extended periods. This is aligned with covering an extended time. We focus on liquidity pool actions, such as liquidity removal, addition, and token burns. The dataset has 15 unique token and liquidity actions from various decentralized exchanges (DEXs), such as Raydium[2] and Jupiter[3]. As rug pull pattern identification is complex, relying on raw transaction data alone is insufficient without a complete view of token liquidity activities. This includes withdrawn liquidity over time or sudden removals. For example, a token may appear normal if its liquidity withdrawal is gradual and may lead to a conclusion that it is not suspicious. Examining the complete history of token activities is required. Thus, our method derives the pattern of suspected rug pull tokens instead of relying solely on raw data. In the following sections, we describe our method for data source selection, selection process, and deriving data from historical transactions and DEX liquidity pool actions.

### 4.1 Time-frame Parameters

We defined three parameters to standardize liquidity action time-frame calculations. The *cutoff_date* is set to November 1, 2024, which is the end date for all of the data temporal calculations. This is important as we excluded token and pool activities after November 01, 2024, to prevent misclassification of token inactivity status. The *inactive_tokens* parameter identifies tokens with no user interactions, such as trading on DEXs after a specific liquidity removal action, such as *RemoveLiquidity*. Also, the *start_date* and *end_date* were used to bound the data retrieval dates, which range from February 12, 2021, to November 1, 2024.

### 4.2 Liquidity Additions and Removals

Our dataset includes 15 unique liquidity actions, such as *deposit*, *addLiquidity*, *withdraw*, and *removeLiquidity*. We implemented two primary `Common Table Expressions (CTEs)`: *RecentLiquidityAdds* and *RecentLiquidityRemoves*, which aggregate liquidity pool transaction actions for the timeframe of *start_date* and *end_date*. CTEs temporarily store query results for further processing [12]. They improve the efficiency of data retrieval by excluding the attributes that are used to derive the data but are not included in the final output.

The *RecentLiquidityAdds* aggregates all liquidity addition actions, including *deposit*, *addLiquidityOneSide*, and *bootstrapLiquidity*. For each *LIQUIDITY_POOL_ADDRESS* and *MINT* pair, we calculate the total added liquidity, the number of liquidity addition transactions, and the timestamp of the most recent addition and removal. *MINT* is the token represented as a public key on the blockchain, which is unique for each token. Each *MINT* has one or more liquidity pools. Similarly, the *RecentLiquidityRemoves* aggregates all liquidity but for removal actions, such as *withdraw*, *removeLiquiditySingleSide*, and *removeAllLiquidity*. Each *LIQUIDITY_POOL_ADDRESS* and *MINT* pair is used to find the total removed liquidity, the number of liquidity removal transactions, and the timestamp of the most recent removal.

From the above CTEs, we merge all addition and removal metrics uniquely identifying *LIQUIDITY_POOL_ADDRESS* and the corresponding *MINT*. From this, we derive the *ADD_TO_REMOVE_RATIO*, which measures net liquidity flow by dividing the total added liquidity by the total removed liquidity. Moreover, we capture the most recent token activity timestamp, whether it is from additions or removals.

### 4.3 Pool and Token Inactivity

We created *LatestSwapActivity* to capture the most recent swap transactions for each *MINT*. This records the timestamp of the latest swap and the associated transaction hash including the last wallet address interacted with the *MINT*. This is important as it helps in identifying when the token became inactive. Also, it aids in identifying the last wallet interacted with the token. This could lead to more interesting findings, such as a single wallet conducting repeated rug pulls that were found on Ethereum and reported by [7]. Finally, it enables the identification of when a token becomes inactive.

---

[2]https://raydium.io/
[3]https://jup.ag/

Using the identified patterns above, we assign an inactivity state to each pool. A token is categorized as *Inactive* if the latest swap occurred following a liquidity removal activity and, from there, the user stopped interacting with the token. The last token activity transaction signals the token lifetime, where the birth of a token is the first mint transaction. Conversely, if users continue to trade the token as of the *end_date*, the token is categorized as *Active*.

### 4.4 Dataset Overview

In Table 1, we show the SolRPDS attributes and types. We make the data available at: *https://github.com/DeFiLabX/SolRPDS*, which is provided in two formats, CSV and JSON. The CSV format organizes the data in a tabular structure, and the JSON format presents the same data in a hierarchical structure, which is suitable for use in data analysis pipelines.

Table 3 provides a summary of the dataset, which contains liquidity addition and removal activities on the Solana blockchain from February 12, 2021, to November 1, 2024. The dataset (Table 2) includes a total of 109,668 aggregated liquidity addition and removal transactions, derived from 3.69 billion blockchain transactions.

*Total Added Liquidity* and *Total Removed Liquidity* represent the cumulative amounts of liquidity added to and removed from liquidity pools, respectively. The mean total added liquidity is $4.99 \times 10^{13}$, with a standard deviation of $1.05 \times 10^{16}$, which represents a considerable variability in liquidity transactions. The minimum added liquidity is 1.0. This was intentional to exclude *MINT* with no liquidity added, which does not add significance from including these tokens. In contrast, the maximum is $3.45 \times 10^{18}$, as there are significant liquidity movements.

The *# of Liquidity Adds* and *# of Liquidity Removes* denote the number of liquidity addition and removal transactions per liquidity pool. On average, there are approximately 1,485 liquidity addition transactions and 1,027 liquidity removal transactions, with standard deviations of $8.11 \times 10^4$ and $8 \times 10^4$, respectively. This high variability suggests that some pools experience a large number of transactions while others remain relatively inactive.

The *Add to Remove Ratio* measures the balance between liquidity inflows and outflows by dividing the total added liquidity by the total removed liquidity. The mean ratio is $6.88 \times 10^4$, with a standard deviation of $8.98 \times 10^6$, and ranges from $1.8 \times 10^{-17}$ to $1.99 \times 10^9$. This shows diverse liquidity dynamics across different pools, with some pools having overwhelmingly higher additions compared to removals and vice versa.

## 5 SolRPDS ANALYSIS AND OBSERVATIONS

This section analyzes the behavior of liquidity pools and shows rug pull patterns. We also present the utility of SolRPDS through machine learning classification experiments for token activity.

**Table 2: Tokens and liquidity pools summary**

| Metric | Count |
| --- | --- |
| Unique Tokens | 33,746 |
| Unique Liquidity Pools | 63,520 |
| Active Tokens | 11,551 |
| Inactive Tokens | 22,195 |

### 5.1 Liquidity Patterns in Rug Pulls

We treat inactivity as a signal for suspicious behavior and not as a definitive evidence for rug pull. Our analysis shows distinct differences between active and inactive pools for removals. Active pools exhibit significant removal actions, with an average of over 80, indicating frequent trading and active pool management associated with legitimate tokens. In contrast, inactive pools display far fewer removals, averaging around 13, with many having only one or very few. This suggests that inactive pools may be abandoned or involved in bots conducting rug pulls. In these cases, funds are added when the token is minted but are suddenly removed once trading starts, generating profit for the entity behind the rug pull. Furthermore, when additions exceed 1000, inactive pools show little to no subsequent activity, whereas active pools remain engaged. These patterns are illustrated in Figure 1.

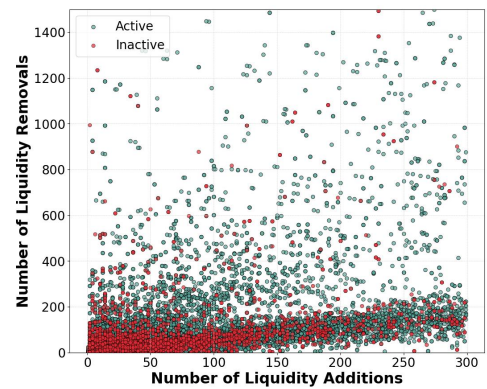

**Figure 1: Comparison of liquidity addition and removals**

### 5.2 Active vs. Inactive Token Trends

Our temporal analysis of the dataset shows a consistent increase in both active and inactive tokens over the examined years. Inactive tokens exhibit significant expansion, particularly in 2023 and 2024. This suggests that the number of projects that have experienced rug pulls or have been abandoned is growing. At the same time, active tokens experienced substantial growth in 2024, indicating increased engagement and ecosystem activity. These dual trends, shown in Figure 2, suggest that while the blockchain ecosystem is attracting more projects, it is also witnessing a rising number of tokens that are becoming inactive.

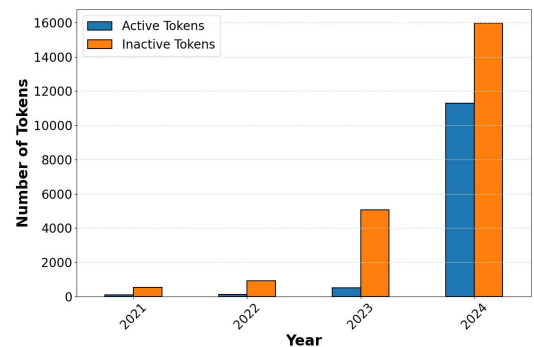

**Figure 2: Yearly distribution of active and inactive tokens**

Table 3: SolRPDS main attributes summary

| Metric | Total Added Liquidity | Total Removed Liquidity | # of Liquidity Adds | # of Liquidity Removes | Add to Remove Ratio |
|---|---|---|---|---|---|
| Count | $1.10 \times 10^5$ | $1.10 \times 10^5$ | $1.10 \times 10^5$ | $1.10 \times 10^5$ | $1.10 \times 10^5$ |
| Mean | $4.99 \times 10^{13}$ | $1.55 \times 10^{14}$ | $1.49 \times 10^3$ | $1.03 \times 10^3$ | $6.88 \times 10^4$ |
| Std | $1.05 \times 10^{16}$ | $2.24 \times 10^{16}$ | $8.11 \times 10^4$ | $8.00 \times 10^4$ | $8.98 \times 10^6$ |
| Min | $1.00 \times 10^{-9}$ | $1.00 \times 10^0$ | $1.00 \times 10^0$ | $1.00 \times 10^0$ | $1.80 \times 10^{-17}$ |
| 25% | $5.40 \times 10^1$ | $1.07 \times 10^2$ | $2.00 \times 10^0$ | $2.00 \times 10^0$ | $4.63 \times 10^{-1}$ |
| 50% | $7.11 \times 10^4$ | $1.11 \times 10^5$ | $5.00 \times 10^0$ | $4.00 \times 10^0$ | $9.16 \times 10^{-1}$ |
| 75% | $4.86 \times 10^6$ | $1.02 \times 10^7$ | $1.90 \times 10^1$ | $1.40 \times 10^1$ | $1.13 \times 10^0$ |
| Max | $3.45 \times 10^{18}$ | $5.57 \times 10^{18}$ | $1.40 \times 10^7$ | $1.41 \times 10^7$ | $1.99 \times 10^9$ |

## 5.3 Comparison of Active and Inactive Pools

A comparative analysis of active and inactive pools (Figure 4) shows significant trends. Active pools that support active tokens are more prevalent than inactive pools. Active tokens maintain consistent trading volumes and balanced liquidity metrics. Conversely, the number of inactive pools is closely aligned with the number of inactive tokens. This suggests that once a pool becomes inactive, users stop interacting with its associated token, which supports the existence of a correlation between pool inactivity and rug pull patterns. These two categories emphasize the importance of monitoring liquidity metrics to detect suspicious activity.

Table 4: Evaluation of token activity state using different ML classifiers

| Model | AUC | ACC | F1 | Prec | Recall | MCC |
|---|---|---|---|---|---|---|
| Random Forest | 0.987 | 0.974 | 0.974 | 0.976 | 0.974 | 0.935 |
| kNN | 0.806 | 0.755 | 0.660 | 0.816 | 0.755 | 0.181 |
| AdaBoost | 0.984 | 0.976 | 0.977 | 0.978 | 0.976 | 0.942 |
| Logistic Regression | 0.877 | 0.898 | 0.889 | 0.911 | 0.898 | 0.728 |
| SVM | 0.748 | 0.866 | 0.858 | 0.863 | 0.866 | 0.626 |
| Neural Network | 0.995 | 0.901 | 0.892 | 0.912 | 0.901 | 0.735 |

## 5.4 Temporal Dynamics of Fraudulent Liquidity Pools

The cumulative fraction of inactive tokens in relation to their duration highlights a remarkable trend. Roughly, 75% of the inactive tokens have a duration of less than one day (Figure 3). This duration is calculated as the time difference between when the token is minted (the creation of the token on the blockchain) and the last user's interactions. This is aligned with [7] findings, where such short-lived tokens are referred to as 1-day tokens. Tokens remain active for a very short period before becoming inactive. The short timeline of activity, where liquidity is added and then withdrawn, changes the token status from active to inactive. Tokens are quickly removed from the market, which leaves limited time for user interaction, such as selling the token to mitigate financial losses.

## 5.5 Training and Testing Classifiers

We have conducted several experiments to demonstrate the utility of the SolRPDS dataset for future research and industry use. Our experiments aim to classify tokens into active or inactive based on the token's liquidity activities. We use six machine learning classification algorithms, shown in table 4, implemented using Python 3 and the scikit-learn library [24]. We sampled two years of the data in SolRPDS to train and test the six algorithms. Each subset consists of a year of data. Then, it was used to predict the tokens' status, classifying them as either active or inactive. SolRPDS provides additional insights beyond token activity classification. However, we focus on the binary classification of token status for simplicity.

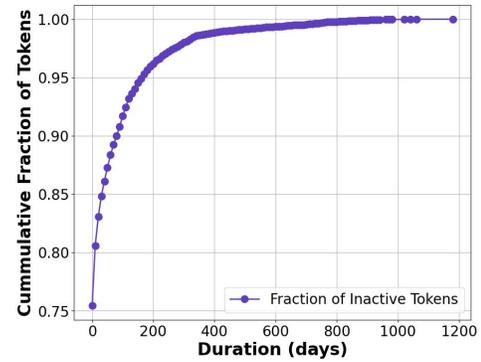

Figure 3: Cumulative fraction of inactive tokens distribution by duration in days

Models were trained using scikit-learn with default settings on 2022 data for training and 2021 data for testing, which covers token and liquidity pool activities. We use attributes such as total added liquidity, total removed liquidity, and add-to-remove ratio, which are shown in Table 1. We computed feature rankings using information gain and excluded features that were less relevant to the model's predictions.

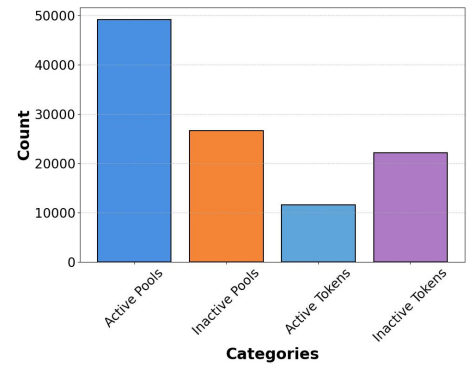

Figure 4: Active vs inactive pools and tokens.

Table 4 shows the results of the classifiers. AdaBoost, an ensemble tree-based algorithm, achieved the highest accuracy of 97.6%, followed by Random Forest (RF) at 97.4%. In contrast, k-nearest neighbors (kNN) was the least accurate, achieving only 75.5%. These highlight the effectiveness of ensemble methods like AdaBoost and RF in capturing token activity patterns compared to simpler models like kNN. Feature importance ranking shows that the number of liquidity removals and additions contributed most to model performance.

The results also show SolRPDS's potential to uncover token activity patterns. Token inactivity classification is a use case of the dataset, and future research may use the data to discover other new pattern insights.

## 6 FUTURE WORK

SolRPDS provides a foundation for further research in rug pull detection and mitigation. It also allows the development and testing of new methods of detection that address the challenges of identifying rug pull patterns in DeFi. Specifically, the annotations of the dataset, such as inactivity periods, liquidity activities, and withdrawn token amounts, can serve as inputs for machine learning models and heuristic-based approaches.

The dataset will also contribute to a better understanding of the behavior of liquidity pools, thereby informing automated detection systems for real-time identification of suspicious activities. Therefore, it reduces financial losses and preserves trust in DeFi platforms. This public dataset will further promote collaboration among researchers and industries to benchmark existing methods and propose novel and robust algorithms for detecting rug pulls.

Future work could use data-driven techniques such as training advanced deep learning models to identify rug pulls based on token inactivity, liquidity shifts, and token price changes. Researchers also may compare rug pull in Solana with other blockchains, such as Ethereum, to derive insights into key differences. On-chain detection methods can be tested against historical events from the dataset to improve real-time detection. Finally, examining correlations between token attributes, liquidity patterns, and token activity may also guide better governance frameworks for DeFi platforms.

## 7 CONCLUSION

This paper presents SolRPDS, the first public dataset designed for rug pull identification on the Solana blockchain. It consists of raw and derived attributes from 2021 to 2024 based on 3.69 billion transactions. We identify 62,895 suspicious liquidity pools along with 22,195 tokens that indicate rug pull patterns. The dataset aims to help the industry and support future research development of robust detection and mitigation methods by simplifying blockchain data and making it publicly available. SolRPDS aims to contribute to a better understanding of rug pulls on Solana for safeguarding DeFi platforms and sustaining users' trust.

## ACKNOWLEDGEMENT

This material is based upon work supported by the National Science Foundation under Grant #2020071. Any opinions, findings, and conclusions or recommendations expressed in this material are those of the authors and do not necessarily reflect the views of the National Science Foundation.